\documentclass[preprint,12pt]{elsarticle}

\usepackage{amsmath}
\usepackage{graphicx}
\usepackage{siunitx}
\usepackage{booktabs}
\usepackage{hyperref}

\usepackage{etoolbox}
\makeatletter
\def\ps@pprintTitle{%
 \let\@oddhead\@empty
 \let\@evenhead\@empty
 \def\@oddfoot{\footnotesize\itshape Preprint. Submitted to Applied Radiation and Isotopes.\hfill\today}%
 \let\@evenfoot\@oddfoot}
\makeatother

\begin{document}

\begin{frontmatter}

\title{Characterisation and Monte Carlo validation of a compact AmBe neutron irradiation facility providing fast and thermal neutron fields for detector development}

\author{A.J.\,Bevan}
\author{I.\,Dawson\corref{cor1}}
\ead{i.dawson@qmul.ac.uk}
\cortext[cor1]{Corresponding author}
\address{School of Physical and Chemical Sciences, Queen Mary University of London, UK}

\begin{abstract}
We report the design, commissioning and benchmarking of a compact AmBe-based neutron irradiation facility capable of providing both fast and thermal neutron dominated fields through multiple detector positions within a moderator assembly.

Detailed radiation transport simulations using the FLUKA Monte Carlo code were performed to model the radiation environment at different detector positions. 
The inclusion of a single-crystal CVD diamond neutron detector in the simulations enabled direct comparison with experimental measurements, 
providing confidence the radiation fields are well understood. 
The simulations also provided a detailed breakdown of energy deposition mechanisms in the diamond sensors, including nuclear recoil, 
neutron capture reactions and secondary proton production from surrounding materials, highlighting the influence of detector housing materials on the local radiation environment and detector response.

The facility provides a practical and accessible platform for neutron detector development and benchmarking in typical university laboratories, 
with dose rates outside the facility below typical natural background levels.
\end{abstract}

\begin{keyword}
Neutron irradiation facility \sep AmBe neutron source \sep FLUKA simulation \sep neutron detector benchmarking \sep radiation transport modelling
\end{keyword}

\end{frontmatter}

\section{Introduction}

The development, characterisation and calibration of neutron radiation detectors require controlled neutron irradiation environments in which the detector response can be studied under well-defined conditions. 
Such facilities are essential for application areas including nuclear instrumentation, radiation protection, fusion diagnostics and reactor monitoring.

Access to neutron irradiation facilities is often limited to research reactors or accelerator-based sources, which, although providing high fluxes and well-characterised spectra, 
are not always readily accessible for routine laboratory use. 
Compact radioisotope-based sources, such as $^{241}$AmBe, offer a practical solution for detector research and development. 
These sources provide a continuous neutron emission spectrum extending up to $\sim$10\,MeV and, when combined with suitable moderator assemblies, 
can produce neutron fields with a range of spectral characteristics.
AmBe sources are particularly attractive because their radiation fields arise from well-defined nuclear processes, 
including the $^{241}$Am($\alpha$,n)$^9$Be reaction and the associated 4.44\,MeV gamma emission from excited $^{12}$C. 

For detector development the AmBe source gamma emission is an advantage. 
In real-world applications, neutron fields are typically accompanied by gamma backgrounds, which means detector performance should be evaluated under such conditions. 
Interpretation of the measured response, however, still requires accurate simulation, as neutron and gamma induced contributions cannot be uniquely separated from measurement alone. 
The combined use of simulation and experiment enables a physically meaningful benchmarking approach for a given application.

For detector development, it is advantageous to have compact facilities capable of providing both fast and thermal neutron fields within a single experimental configuration. 
Such facilities allow systematic studies of detector response under different spectral conditions while remaining compatible with standard laboratory radiation protection requirements.
The facility is intended for detector response studies and benchmarking rather than radiation damage testing.

In this work we report the design, commissioning and characterisation of a compact AmBe-based neutron irradiation facility for detector research and development. 
Detailed radiation transport simulations using the FLUKA Monte Carlo code were performed to model the neutron and gamma fields, 
and the predicted detector response was benchmarked against experimental measurements obtained with a commercial diamond detector. 
The simulations also provide insight into the interaction mechanisms contributing to detector response, including nuclear recoil, 
neutron capture and secondary particle production originating in the detector packaging material.

While neutron-field characterisation is often performed using activation foils, the present approach benchmarks the facility using detector response measurements, 
providing a more relevant validation for detector development and naturally incorporating local material and packaging effects that are system dependent.

The results demonstrate that the radiation environment within the facility is well understood and that the combined simulation and experimental approach provides a robust framework for neutron detector benchmarking in a compact laboratory setting with a minimal radiological footprint.

\section{Facility design}

\subsection{Overview}
\label{sec:fac_overview}

\begin{figure}[htbp]
    \centering
    \includegraphics[width=0.5\linewidth]{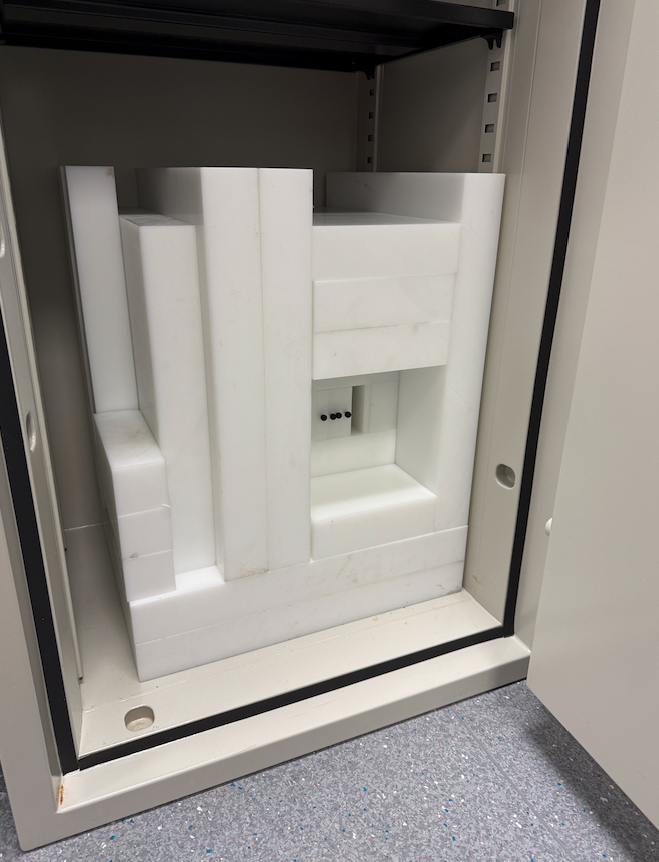}
    \caption{The source sits permanently within a HDPE moderator block, which is surrounded by further HDPE for shielding and enclosed within a fire-rated steel safe to comply with safety and security requirements.}
    \label{fig:safe}
\end{figure}

The facility consists of a compact high density polyethylene (HDPE) moderator assembly containing a permanently housed AmBe source, 
surrounded by additional HDPE for shielding, with everything enclosed within a fire-rated steel safe as shown in Figure~\ref{fig:safe}.
The design allows detectors to be inserted at several defined positions within the moderator block, 
enabling measurements in neutron fields ranging from fast neutron dominated to strongly moderated conditions. 

\subsection{The AmBe source}
\label{sec:ambe}

An $^{241}$Am-Be (AmBe) source was used, with a certified neutron emission rate of 18500\,$\pm$\,320 neutrons per second.\footnote{Measured and certified by the UK's National Physical Laboratory, November 2025.}
The source is cylindrical, with outer dimensions of 7.8\,mm diameter and 10\,mm height, comprising an AmBe pellet encapsulated by stainless steel.

\begin{figure}[htbp]
    \centering
    \includegraphics[width=0.48\linewidth]{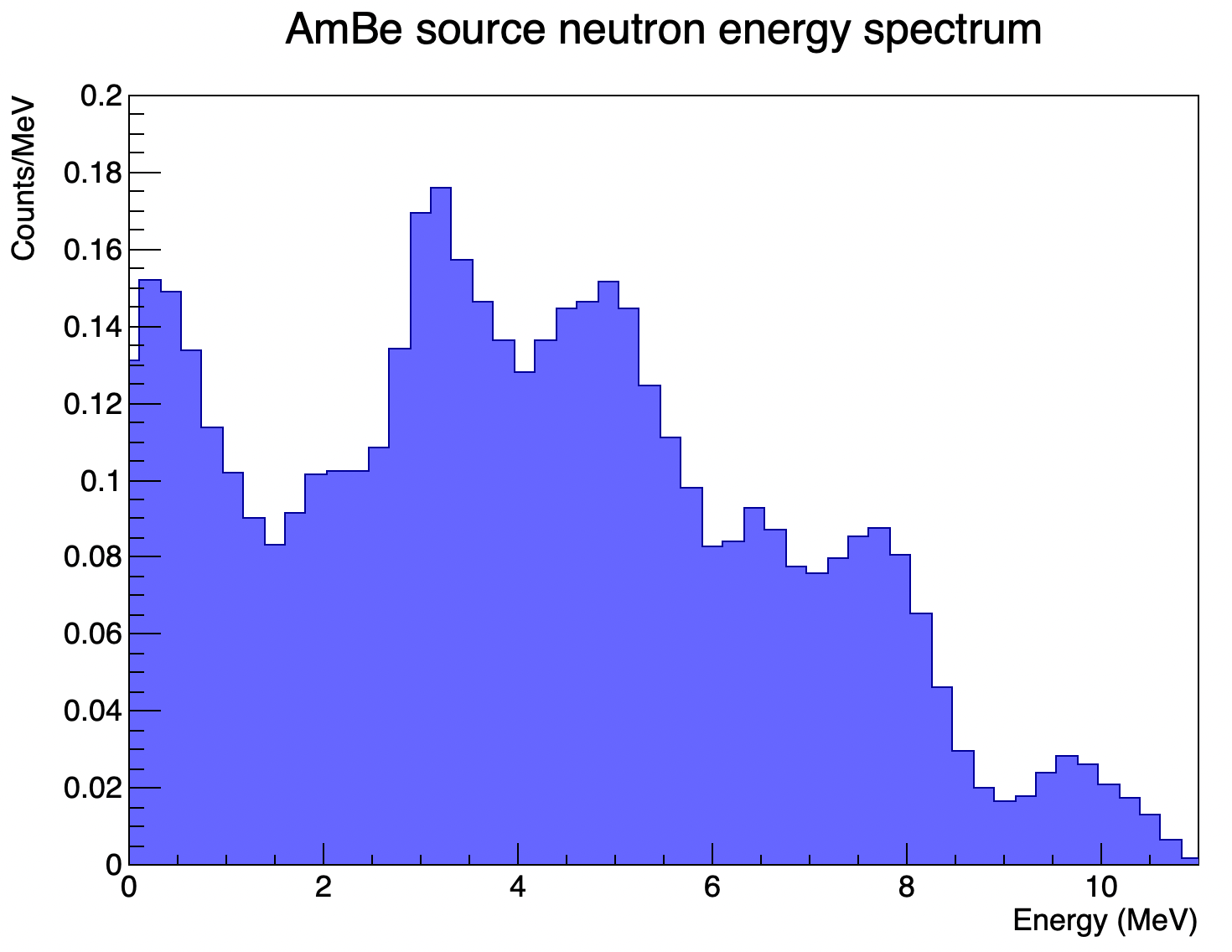}
    \hfill
    \includegraphics[width=0.49\linewidth]{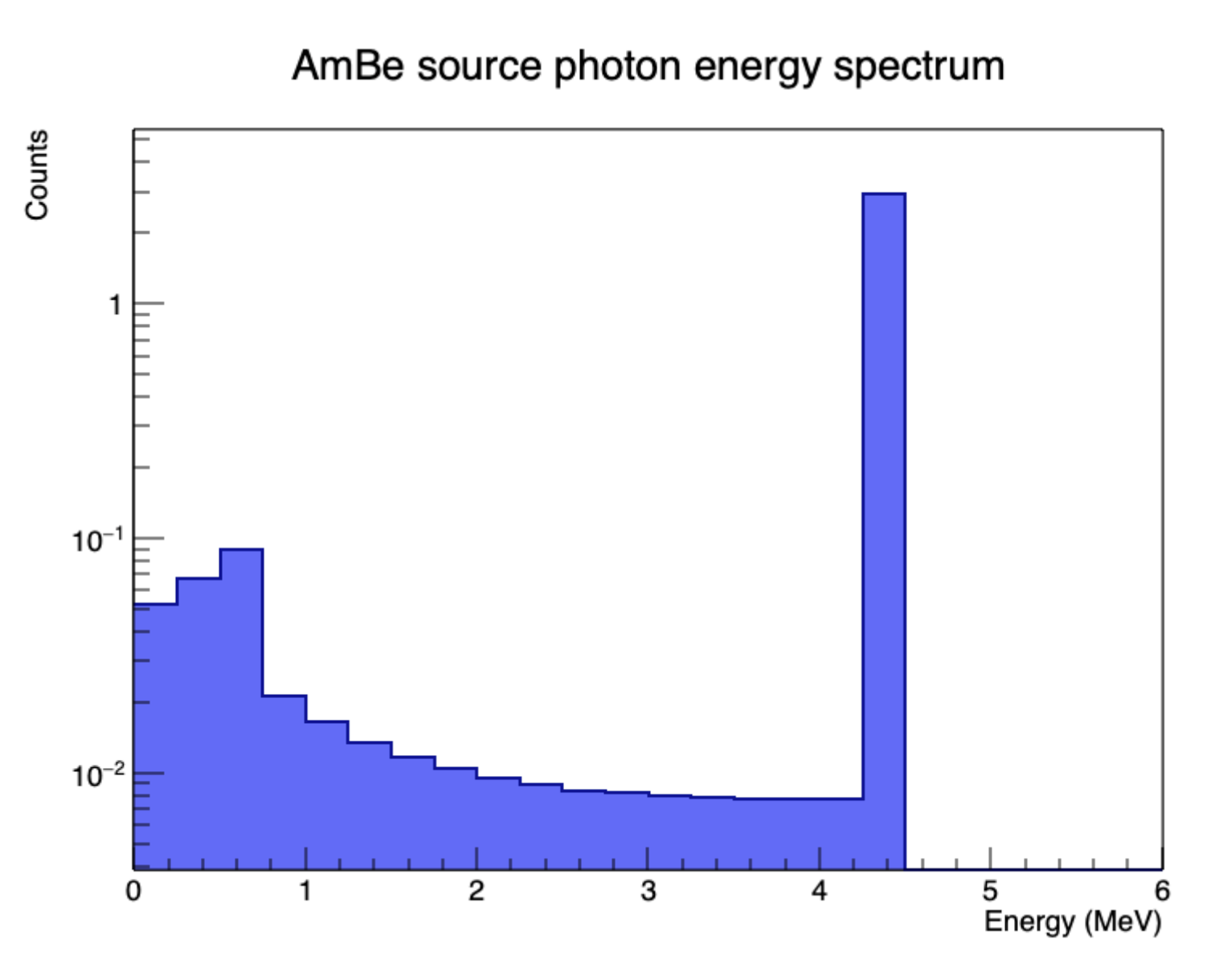}
    \caption{Left: neutron energy spectrum from an AmBe source. 
    Right: corresponding gamma emission spectrum, dominated by the 4.44\,MeV line with a smaller contribution from lower-energy secondary photons created within the source material.}
    \label{fig:ambe_spectra}
\end{figure}

Figure~\ref{fig:ambe_spectra} shows the neutron and photon energy spectra produced by the source. 
The neutron spectrum follows the ISO~8529-1:2021 reference spectrum \cite{ISO-2021}.
The photon spectrum is dominated by the 4.44\,MeV gamma-ray emission, 
with additional lower-energy photons arising from secondary interactions within the source encapsulation material.

\subsection{HDPE moderator setup}
\label{sec:HDPEsetup}

\begin{figure}[htbp]
    \centering
    \includegraphics[width=0.9\linewidth]{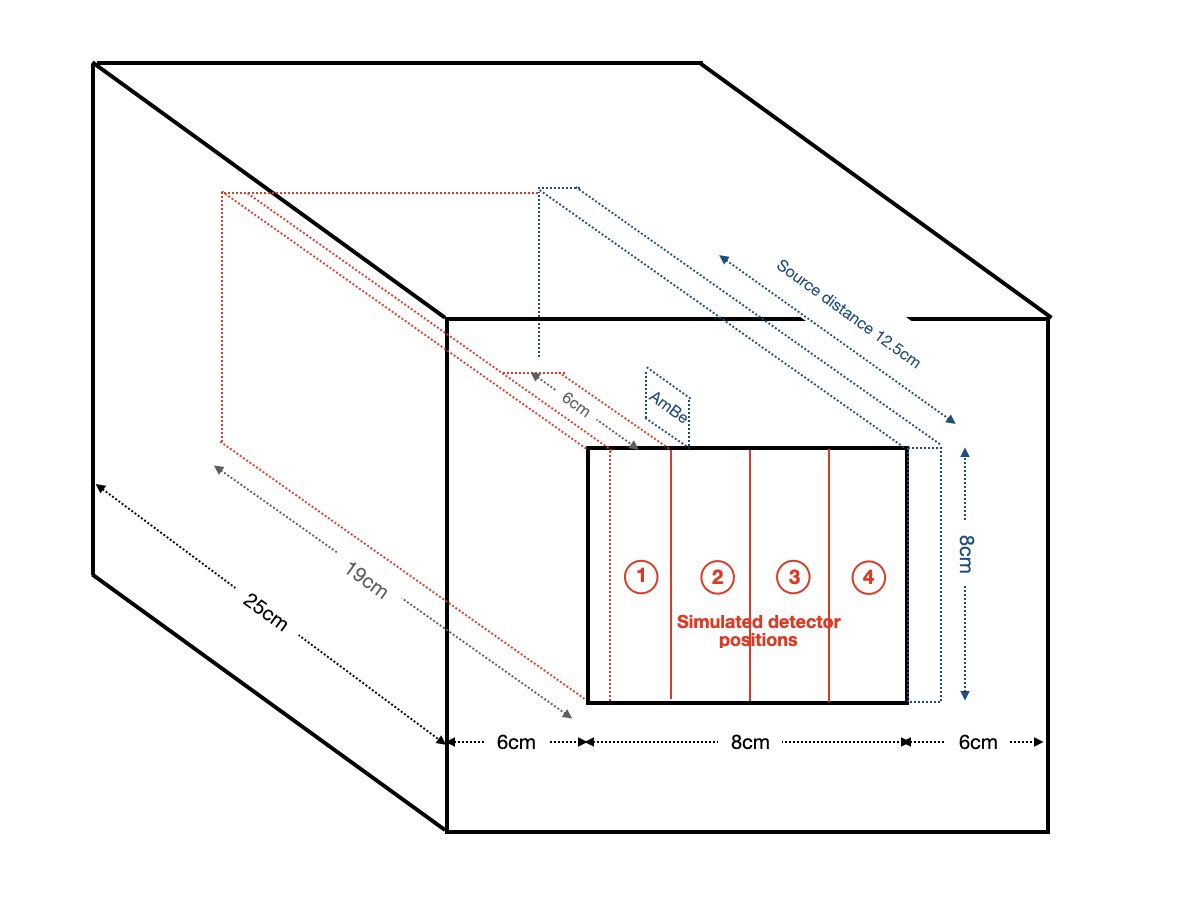}
    \caption{Moderator block with removable HDPE fillers that allow variable thicknesses of moderator material between the AmBe source and the device under test. 
    The relative dimensions shown are schematic and not to scale.}
    \label{fig:mod_block}
\end{figure}

To thermalise the fast neutrons emitted by the AmBe source, a compact system was constructed from HDPE, 
the details of which are shown in Figure~\ref{fig:mod_block}. 
HDPE was chosen for its high hydrogen content\footnote{HDPE is a polymer composed of repeating –CH$_2$– units.} and effective moderating capability. 
The outer dimensions of the block are 20\,cm\,$\times$\,20\,cm\,$\times$\,25\,cm, and the density of the HDPE material was measured to be 0.963\,g/cm\textsuperscript{3}.

The inner volume, into which modular inserts (or ``fillers'') are placed, is 8\,cm\,$\times$\,8\,cm\,$\times$\,19\,cm. 
The filler thicknesses are 1\,cm and 2\,cm which can be repositioned or removed to vary the amount of moderator material between the source and the detector. 
This modularity allows control of the neutron energy spectrum incident on the detector, 
supporting systematic studies across a range of field conditions, from fast-dominated to thermal-dominated.

In addition, bespoke HDPE detector holders were designed to position the detectors reproducibly within the selected moderator locations. 
These holders ensured stable mechanical placement and a well-defined detector position relative to the AmBe source, 
improving both measurement repeatability and consistency with the simulated geometry.

\section{Monte Carlo modelling of the neutron irradiation facility}
\label{sec:fluka}

FLUKA is a general-purpose Monte Carlo code developed at CERN for the simulation of particle transport and interactions over a wide energy range, 
from thermal neutrons to multi-TeV hadrons and leptons. 
It provides detailed models for electromagnetic and hadronic processes and is widely used in radiation shielding, dosimetry, 
medical physics and detector design~\cite{Ahdida2022, Battistoni2015}.

In this work, FLUKA version~4.5.1 was used to simulate neutron transport, moderation, scattering and energy deposition in the experimental configuration. 
The full geometry of the facility was implemented in the model, including the HDPE moderator block described in Section~\ref{sec:HDPEsetup}. 
The moderator was assigned a density of 0.963\,g\,cm$^{-3}$ and chemical composition corresponding to polyethylene (C$_2$H$_4$)$_{\text n}$.
The geometry also included the HDPE detector holders used in the experimental configuration, 
ensuring the detector position relative to the source and moderator was modelled consistently with the measurements.

Neutron and gamma emissions from the $^{241}$AmBe source were implemented using the spectra described in Section~\ref{sec:ambe}. 
Primary particles were generated isotropically within the AmBe pellet using the FLUKA \texttt{source.f} routine, 
and transported through the moderator and detector materials. 
Typically $5\times10^8$ source neutrons and gammas were generated separately for each configuration.

Average neutron fluxes at the four detector locations were scored using the \texttt{USRTRACK} estimator, 
while energy deposition in a diamond sensor was obtained using \texttt{EVENTBIN} on an event-by-event basis. 
The neutron fluxes were evaluated both with and without the detector assembly present, 
allowing the external neutron flux to be distinguished from local perturbations caused by detector packaging materials. 
It is well known that detector packaging and surrounding materials can locally perturb the neutron field and therefore influence the energy deposited within the diamond sensor.

The FLUKA \texttt{PRECISIOn} defaults were used to enable detailed modelling of nuclear and electromagnetic interactions, including evaporation processes and ion transport. Neutron interactions are treated using point-wise cross sections from the JEFF-3.3 evaluated nuclear data library~\cite{JEFF33}. 
Transport thresholds were set to 10\,keV for all particles except neutrons, 
which were transported down to \SI{1e-14}{GeV} to ensure accurate treatment of thermal neutron interactions.

\section{Simulated neutron and gamma fluxes at detector positions}
\label{sec:sim_fluxes}

\begin{figure}[htbp]
    \centering
    \includegraphics[width=0.9\linewidth]{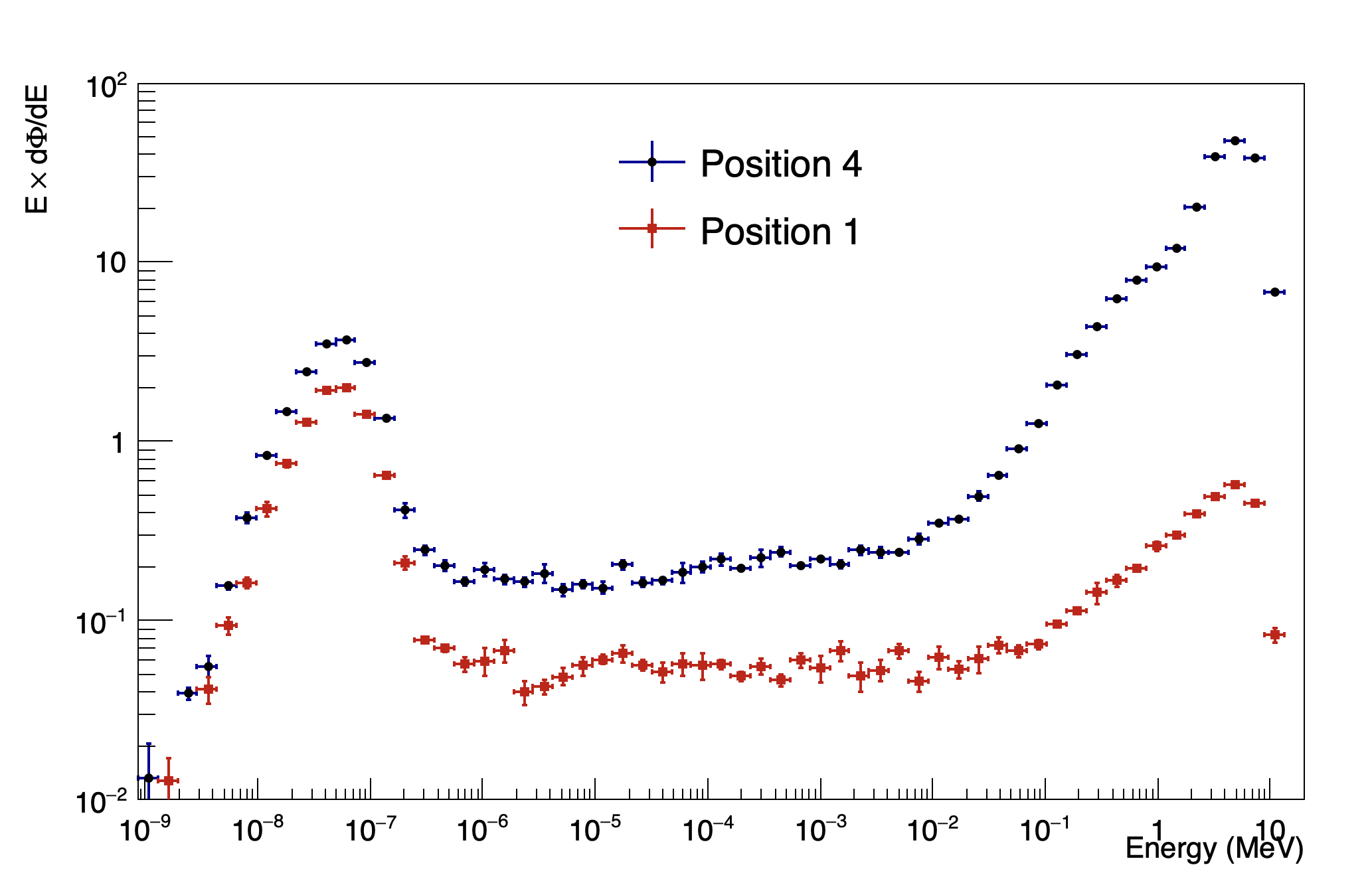}
    \caption{Simulated neutron energy spectra at detector positions~1 and~4, illustrating the effect of the HDPE moderator on the AmBe neutron field. 
    Position~4 is dominated by fast neutrons, whereas position~1 shows that, while the total flux is reduced following moderation, the relative thermal component is significantly enhanced.}
    \label{fig:nspec}
\end{figure}

The four detector positions shown in Figure~\ref{fig:mod_block} provide a controlled progression from fast-neutron dominated to strongly moderated neutron fields, 
making the facility suitable for benchmarking detectors under different spectral conditions. 
The neutron fluxes at the detector positions are summarised in Table~\ref{tab:neutron_fluxes}.
At position~4, located closest to the AmBe source, the neutron field is predominantly fast, with only $\sim$\,8\% thermal neutrons. 
In contrast, at position~1, the furthest location from the source and separated by approximately 6\,cm of moderator, 
the thermal neutron fraction increases to more than 65\% of the total neutron flux.
Figure~\ref{fig:nspec}  shows the corresponding neutron energy spectra for detector positions~1 and~4, 
illustrating the effect of the HDPE moderator on the AmBe neutron field. 

\begin{table}[htbp]
\centering
\caption{Simulated mean neutron fluxes (cm$^{-2}$\,s$^{-1}$) for the four detector locations within the moderator assembly. 
The fluxes are grouped as thermal (E\,$<$\,0.5\,eV), epithermal/intermediate (0.5\,eV\,$\leq$\,E\,$<$\,100\,keV), and fast (E\,$\geq$\,100\,keV). 
The values have been normalised to the AmBe source emission rate of 18{,}500\,n/s.
Uncertainties are dominated by the $\sim$\,2\% uncertainty on the neutron emission normalisation.}
\begin{tabular}{ccccc}
\cmidrule(lr){1-5}
 Position & Total & Thermal & Epithermal & Fast \\
\midrule
 1  & 107  & 70  & 12  & 25  \\
 2 & 183  & 103 & 25  & 55  \\
 3 & 345  & 132 & 47  & 165 \\
 4 & 1683 & 134 & 63  & 1485 \\
\bottomrule
\end{tabular}
\label{tab:neutron_fluxes}
\end{table}

The modular design of the moderator also allows additional shielding materials to be introduced in order to modify the mixed radiation field. 
As an example, simulations were performed for position~1 in which the HDPE in positions~3 and~4 was replaced by 4\,cm of lead, leaving 2\,cm of HDPE between the source and detector. 
This configuration was investigated as a means of reducing the primary gamma component from the AmBe source while preserving a moderated neutron field. 
As shown in Table~\ref{tab:flux_comp}, the primary photon flux above 1\,keV is reduced from 40.6 to 17.2\,cm$^{-2}$\,s$^{-1}$, 
while the primary photon flux above 1\,MeV is significantly reduced from 26.4 to 10.3\,cm$^{-2}$\,s$^{-1}$. 
This therefore provides an additional configuration with an improved neutron-to-photon ratio for detector response studies.

A further possible configuration is the use of a thin cadmium liner around the irradiation volume to suppress the  thermal neutron component. 
This would allow position~4 to be operated as a predominantly fast neutron field.

\begin{table}[htbp]
\caption{Simulated neutron and gamma fluxes at position~1, comparing the baseline moderator configuration with a configuration in which the HDPE in positions~3 and~4 is replaced by 4\,cm of lead, 
leaving 2\,cm of HDPE between the source and detector. 
The gamma fluxes labelled PI and NI correspond to primary-induced and neutron-induced components, respectively, obtained from separate simulations of AmBe gamma and neutron emissions.
Uncertainties are dominated by the $\sim$\,2\% uncertainty on the neutron emission normalisation.}
\centering
\begin{tabular}{lcc}
 & \multicolumn{2}{c}{Position 1 fluxes (cm$^{-2}$\,s$^{-1}$)} \\
  &  Baseline &  4\,cm\,Pb \\
\midrule
$n_{\scriptscriptstyle \mathrm{total}}$   &  105 &  113 \\
$n_{\scriptscriptstyle \mathrm{thermal}}$   &  68.2 &  67.8 \\
$n_{\scriptscriptstyle \mathrm{fast}}$   &  23.9 & 28.5  \\
$\gamma^{\scriptscriptstyle \mathrm{PI}}_{\scriptscriptstyle \mathrm{>1MeV}}$   &   26.4 &  10.3 \\
$\gamma^{\scriptscriptstyle \mathrm{NI}}_{\scriptscriptstyle \mathrm{>1MeV}}$   &  15.5  &   12.6 \\
$\gamma^{\scriptscriptstyle \mathrm{PI}}_{\scriptscriptstyle \mathrm{>1keV}}$   & 40.6 &   17.2 \\
$\gamma^{\scriptscriptstyle \mathrm{NI}}_{\scriptscriptstyle \mathrm{>1keV}}$   &  25.6 &  21.1  \\
\bottomrule
\end{tabular}
\label{tab:flux_comp}
\end{table}

\section{Experimental setup and detector implementation}
\label{sec:expt}

\subsection{Diamond detector}

A commercial neutron detector was used in this work~\cite{civi_ref}, 
based on a single-crystal chemical vapour deposition (CVD) diamond sensor of thickness 140\,$\mu$m and active area 10\,mm$^2$. 
Fast neutrons are detected primarily via elastic scattering in the diamond, producing nuclear recoils that deposit energy within the sensitive volume. 
Thermal neutrons are detected using a 1.8\,$\mu$m thick $^6$LiF conversion layer deposited near the diamond surface, 
where capture via the $^6$Li(n,$\alpha$)$^3$H reaction produces a 2.06\,MeV $\alpha$ particle and a 2.73\,MeV triton. 
These charged particles enter the diamond and deposit their energy, 
giving rise to a characteristic spectral feature that enables direct identification of the thermal neutron component.

The sensor is integrated into a compact FR4 PCB housing with gold-coated outer surfaces and an SMA connector providing signal readout and bias supply. 
The detector is therefore sensitive to both fast and thermal neutrons, making it well suited for benchmarking the radiation fields within the facility. 
Further details of the detector and its performance are given in~\cite{cividec_perf}.

The choice of this detector was motivated both by its published performance characteristics and by its packaging materials, in particular the use of FR4, which contains hydrogen and trace amounts of boron. 
Such materials are known to influence the local neutron field through additional moderation, absorption and secondary particle production. 
In contrast to low-interaction materials such as aluminium, the FR4 housing is therefore expected to perturb the neutron environment at the sensor location. 
The use of this detector thus provides a representative and practically relevant configuration for studying the combined effects of neutron transport, detector response and material-induced local field perturbations.

\subsection{Signal readout and pulse-height digitisation}

The diamond detector was read out using a Cremat CR-Z-110 charge-sensitive preamplifier~\cite{pre-amp}. 
The resulting signals were processed and pulse-height digitised using a Lynx Multi Channel Analyser (MCA) to produce pulse-height spectra. 
The detector was operated at a DC bias voltage of 120\,V, applied via the input filter of the Cremat amplifier. 
A Tektronix TDS~2022C oscilloscope (200\,MHz bandwidth) was used for real-time monitoring and verification of the detector signals.

\subsection{Energy calibration}

The relationship between MCA channel number and deposited energy was determined in situ using the triton peak from the $^6$Li(n,t)$^4$He reaction (see Figure~\ref{fig:edep_pos1}). 
This feature is well separated from other contributions in both the measured and simulated spectra, 
allowing a reliable determination of its mean position and associated uncertainty.

The mean channel number of the measured peak (783\,$\pm$\,3) was aligned with the mean deposited energy predicted by FLUKA (2.56\,$\pm$\,0.01\,MeV). 
This approach is fully consistent with the experimental configuration and inherently accounts for energy losses in the $^6$LiF conversion layer and surrounding materials prior to energy deposition in the diamond. 
It also reduces reliance on external calibration sources, which may not fully reproduce the detector response under neutron irradiation conditions.

A linear relationship between MCA channel and deposited energy was assumed, with channel zero corresponding to zero deposited energy.

\subsection{Detector model used in the Monte Carlo simulations}

A detailed geometrical model of the CIVIDEC diamond detector was implemented in the FLUKA simulations in order to predict the detector response within the neutron irradiation facility. 
The model included the main structural and functional elements of the packaged detector, rather than representing the sensor as an isolated sensitive volume. 
In particular, the simulation geometry explicitly included the FR4 housing with gold-plated outer layers, active diamond sensor volume with gold electrodes, 
an internal air gap, and a 95\% enriched $^6$LiF neutron conversion layer.

The active diamond region was modelled as a square volume of dimensions $3.2 \times 3.2$\,mm$^2$ and thickness 140\,$\mu$m, consistent with the nominal sensor dimensions. 
The enriched $^6$LiF conversion layer was placed adjacent to the detector and modelled with a thickness of 1.8\,$\mu$m. 
Thin gold electrode layers of thickness 200\,nm were included on both sides of the diamond. 
The surrounding detector structure was represented using FR4, giving total detector dimensions of 44\,mm\,$\times$\,47\,mm\,$\times$\,6.4\,mm. 
Including all packaging material allows its influence on the local neutron field and detector response to be taken into account.

For each simulation, only one detector position was treated as active, while the remaining detector locations were replaced by HDPE.
This allowed the detector response to be evaluated independently at each moderator position without perturbation from additional detector assemblies. 
The detector was placed in an air-filled cavity within the moderator block, reproducing the experimental configuration.

Energy deposition within the active diamond volume was scored on an event-by-event basis using \texttt{EVENTBIN}. 
In addition to the total deposited energy, separate scoring was performed for contributions from heavy ions, tritons, alpha particles, $^7$Li ions, $^{12}$C recoils, protons, electrons and photons. 
This enabled the different physical mechanisms contributing to the detector response to be identified and compared for the various neutron field configurations. 
Neutron and photon fluence spectra within the diamond region were also scored using \texttt{USRTRACK}.

\section{Experimental validation of the radiation field model}

\subsection{Comparison of simulated and measured energy deposition in the diamond sensor}

The measured energy deposition spectrum represents the total detector response to the mixed neutron--gamma field produced by the AmBe source. 
Since the source is calibrated in terms of neutron emission rate, the measured spectrum is normalised to the neutron yield, giving detector response per emitted neutron. 
The simulated mixed field spectrum is constructed by summing the neutron induced response per neutron primary and the gamma-induced response per photon primary, 
weighted by the assumed gamma-to-neutron emission ratio.
In this work, a gamma-to-neutron emission ratio of unity is assumed, although values closer to 0.7 have been reported in the literature~\cite{Ito2020}.
This mixed-field approach also reflects realistic operating conditions in which neutron fields are typically accompanied by gamma backgrounds, 
and therefore provides a more application-relevant validation of detector performance.

\begin{figure}[htbp]
    \centering
    \includegraphics[width=1.0\linewidth]{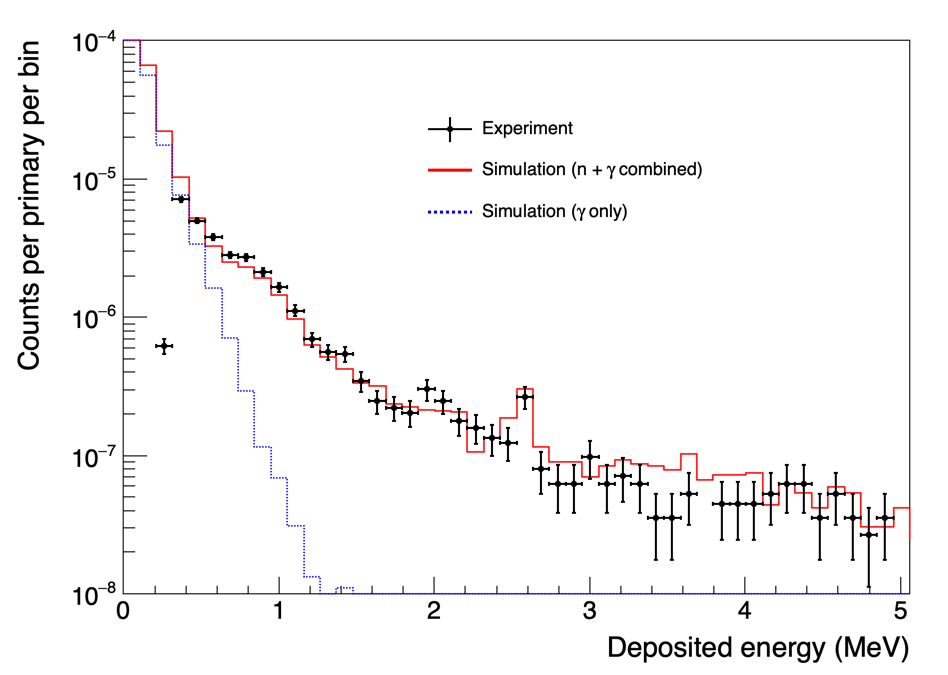}
    \caption{Comparison of simulated and measured deposited-energy spectra in the diamond detector at position~4, corresponding to a fast-neutron-dominated field. 
    The total simulated response (red) is shown along with the photon contribution only (blue). 
    Good agreement between simulation and experiment is observed above $\sim$0.4\,MeV, with the detector response dominated by neutron interactions.}
    \label{fig:edep_pos4}
\end{figure}

\begin{figure}[htbp]
    \centering
    \includegraphics[width=1.0\linewidth]{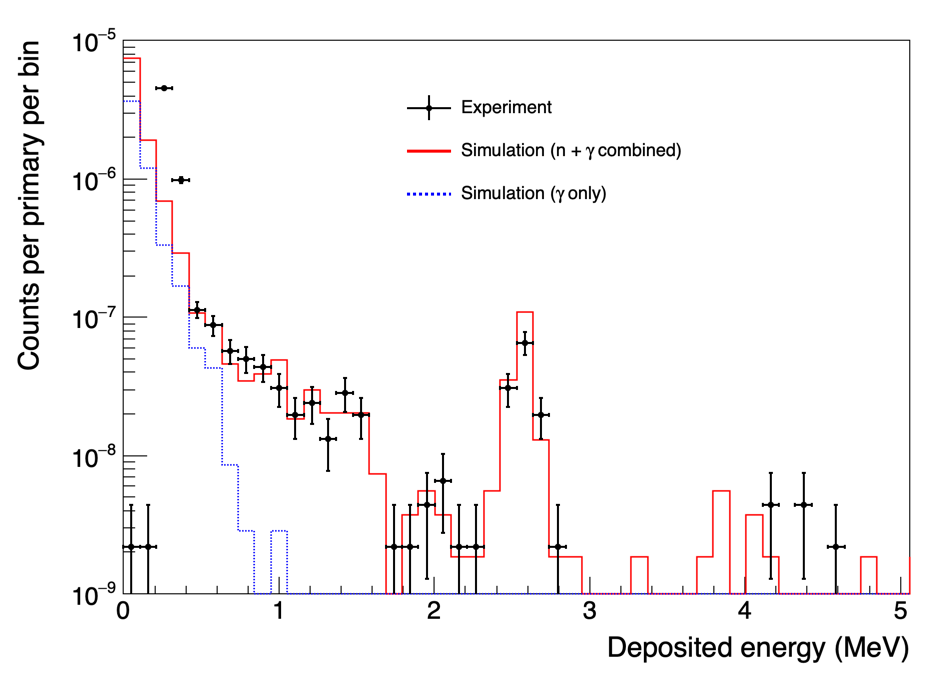}
    \caption{Comparison of simulated and measured deposited-energy spectra in the diamond detector at position~1, corresponding to a strongly moderated neutron field. A prominent peak at $\sim$2.5--2.7\,MeV, arising from the $^6$Li(n,$\alpha$)$^3$H reaction, is clearly reproduced by the simulation. The low-energy region is dominated by gamma-induced contributions in the simulation and by electronic noise effects in the measurement.}
    \label{fig:edep_pos1}
\end{figure}

The comparison between simulated and measured deposited-energy spectra for detector positions~1 and~4 is shown in Figures~\ref{fig:edep_pos4} and~\ref{fig:edep_pos1}. 
In both cases, good agreement is observed between simulation and measurement above approximately 300-400\,keV, both in spectral shape and absolute normalisation. 
This agreement provides strong support for the Monte Carlo description of the neutron and gamma radiation fields within the facility.

At position~4, located closest to the AmBe source, the detector response is dominated by fast neutrons. 
The simulated neutron-induced contribution reproduces both the overall spectral distribution and the high-energy features observed in the measured data, 
while the gamma contribution is confined to low deposited energies and remains a minor component of the total response.

At position~1, where the neutron field is strongly moderated, a pronounced peak is observed at deposited energies around 2.5--2.7\,MeV, 
corresponding to the $^3$H (triton) particle produced in the $^6$Li(n,$\alpha$)$^3$H reaction. 
The simulation accurately reproduces both the position and relative magnitude of this feature, 
demonstrating that the thermal neutron component of the radiation field is well described.

At deposited energies below approximately 300--400\,keV, the agreement between simulation and measurement deteriorates. 
In this region, low-energy neutron and gamma-induced signals are affected by the noise floor, 
threshold behaviour, and pulse-processing characteristics of the data-acquisition system, including the Lynx MCA. 
These effects are not explicitly modelled and are therefore expected to contribute to the observed discrepancies.

\subsection{Contributions from different interaction mechanisms}

\begin{figure}[htbp]
    \centering
    \includegraphics[width=0.8\linewidth]{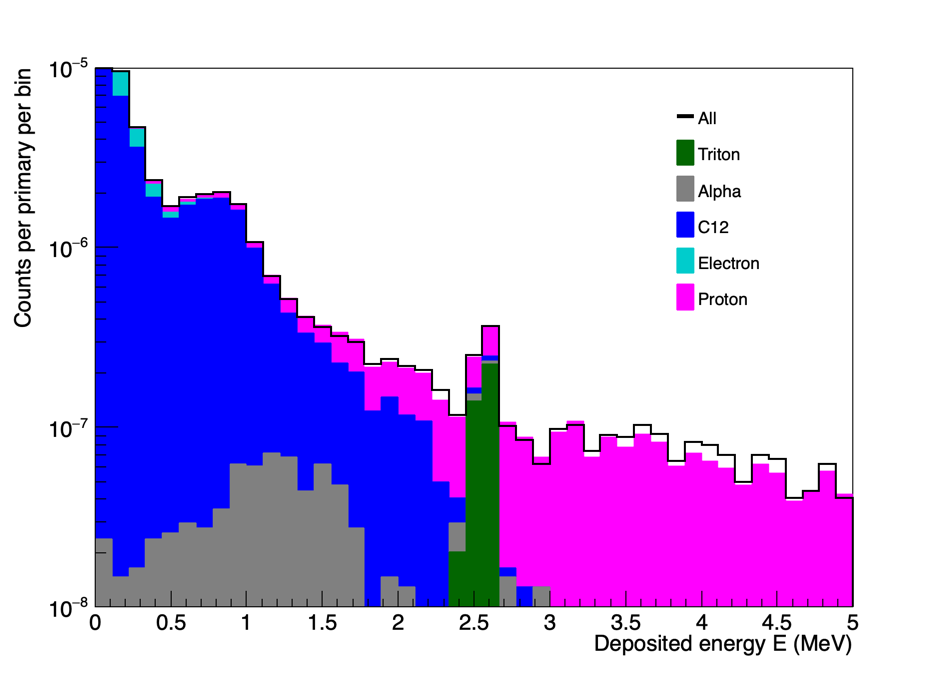}
    \caption{Simulated deposited energy spectra in the diamond detector at position~4, decomposed into contributions from different interaction mechanisms.}
    \label{fig:ncontr_pos4}
\end{figure}

\begin{figure}[htbp]
    \centering
    \includegraphics[width=0.8\linewidth]{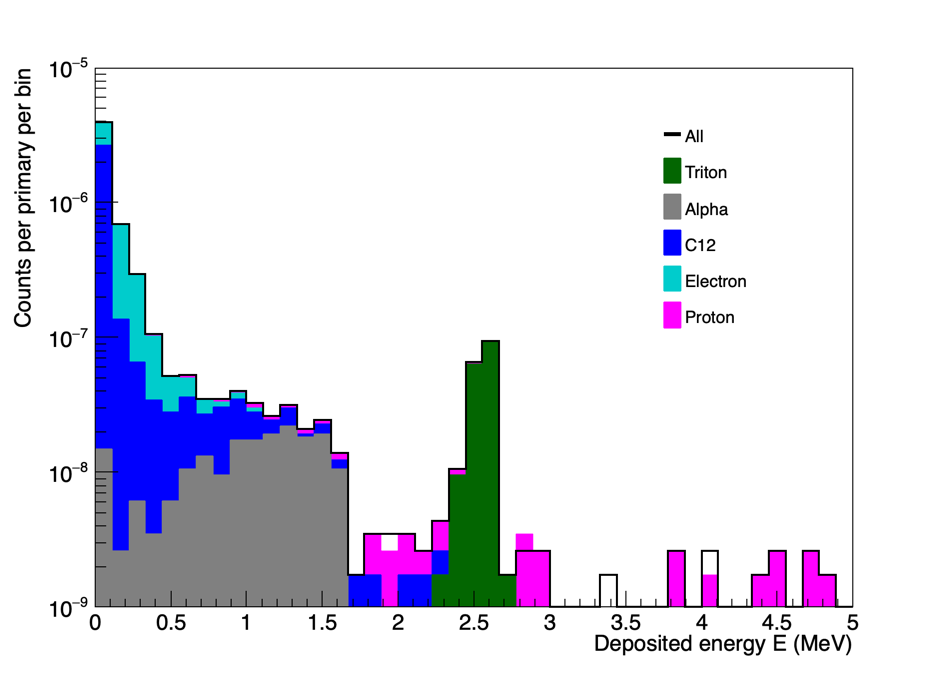}
    \caption{Simulated deposited energy spectra in the diamond detector at position~1, decomposed into contributions from different interaction mechanisms.}
    \label{fig:ncontr_pos1}
\end{figure}

The simulated decomposition of the deposited-energy spectra into individual interaction channels is shown in Figures~\ref{fig:ncontr_pos4} and~\ref{fig:ncontr_pos1} for positions~4 and~1, respectively. 
This breakdown provides physical insight into the origin of the detector response in the different neutron fields.

At position~4, the response is dominated by fast neutron interactions in the diamond and surrounding materials, 
producing a broad distribution of deposited energies through recoil processes and secondary charged particles. 
At position~1, neutron capture in the $^6$LiF conversion layer becomes dominant, giving rise to the characteristic peak associated with emission of triton.

The simulations suggest that detector packaging and surrounding materials contribute to the observed response through secondary particle production, 
including protons and heavier recoils, thereby influencing the local radiation environment at the sensitive volume, as quantified in Section~6.3.

Overall, the detector response can be directly interpreted in terms of specific nuclear interaction channels, including elastic scattering in carbon, neutron capture in $^6$Li, 
and secondary charged-particle production. This reaction-resolved description provides a physically meaningful validation of the radiation field beyond integral flux measurements.

\subsection{Influence of detector materials on the local neutron field}

The detector response described above is calculated using a detailed geometrical model that includes the sensor and its surrounding materials. 
These materials can also locally perturb the radiation field at the sensor location. 
To quantify this effect for neutrons, additional simulations were performed in which the detector assembly was removed and replaced by air, 
allowing comparison between the external neutron flux and the flux in the presence of the detector.

Table~\ref{tab:detector_perturbation} summarises the neutron flux and spectral composition at selected detector positions for the two configurations. 
The results show that the presence of the detector and its packaging leads to a significant modification of the local neutron field at the sensor location. 
In particular, inclusion of the full detector assembly reduces the thermal neutron flux by up to a factor of three. 
This reflects additional scattering, moderation and absorption processes introduced by materials such as FR4, polyethylene and the $^6$LiF conversion layer.

These findings demonstrate that the neutron field experienced by the sensitive detector volume can differ substantially from the external field, 
and that accurate prediction of detector response requires inclusion of the full detector geometry and surrounding materials in the simulation. 
This level of detail is essential for accurate benchmarking of detector performance in realistic measurement conditions and highlights the advantage of detector-based benchmarking approaches, 
which inherently capture local material effects not directly accessible through integral flux measurements alone.

\begin{table}[htbp]
\centering
\caption{Comparison of neutron fluxes (cm$^{-2}$\,s$^{-1}$) at the detector sensor location with and without the detector assembly present.
Thermal and fast neutrons are defined to have energies $<$\,0.5\,eV and $>$\,100\,keV respectively.
Uncertainties on the fluxes are dominated by the $\sim$\,2\% uncertainty on the neutron emission normalisation.}
\label{tab:detector_perturbation}
\begin{tabular}{ccccc}
\hline
Position & Configuration & Total & Thermal & Fast \\
\hline
1 & No detector   & 107 & 69.5 & 24.7 \\
1 & With detector & 63.1 & 22.8 & 25.7 \\
4 & No detector   & 1683 & 134 & 1485 \\
4 & With detector & 1731 & 52.6 & 1572 \\
\hline
\end{tabular}
\end{table}

\subsection{Validation of the neutron irradiation facility}

The agreement between simulated and measured deposited energy spectra across both fast and thermal-neutron dominated configurations demonstrates that the radiation transport model provides an accurate description of the neutron and gamma fields within the facility. 
In particular, the ability of the simulation to reproduce both the broad fast-neutron response and the distinct thermal-neutron capture features confirms that neutron moderation, 
transport and interaction processes are well modelled.

The remaining discrepancies at low deposited energies are attributed primarily to experimental effects associated with detector noise and signal processing, 
rather than deficiencies in the radiation transport model. 
These effects do not significantly affect the interpretation of the detector response in the energy region relevant for neutron detection and benchmarking.

Taken together, the results demonstrate that the combined Monte Carlo modelling and experimental measurements provide a robust validation of the neutron irradiation facility. 
The radiation fields are shown to be well characterised and reproducible across different moderator configurations, 
supporting the use of the facility as a reliable and practical platform for neutron detector development and benchmarking in typical laboratory environments.

\section{Radiological assessment of the facility}
\label{sec:radio}

The AmBe source used in the present work has a certified neutron emission rate of
$18500 \pm 320$ neutrons per second, corresponding to an AmBe source activity of
approximately 300\,MBq. Compared with larger neutron irradiation facilities,
this relatively modest activity represents a practical compromise: it is
sufficient to perform meaningful neutron detector research and development
while remaining compatible with standard laboratory radiation protection
procedures. The AmBe source remains permanently housed within the moderator
assembly and does not require routine handling, thereby minimising personnel
exposure and supporting operation consistent with the ALARP principle.

\begin{figure}[htbp]
    \centering
    \includegraphics[width=0.9\linewidth]{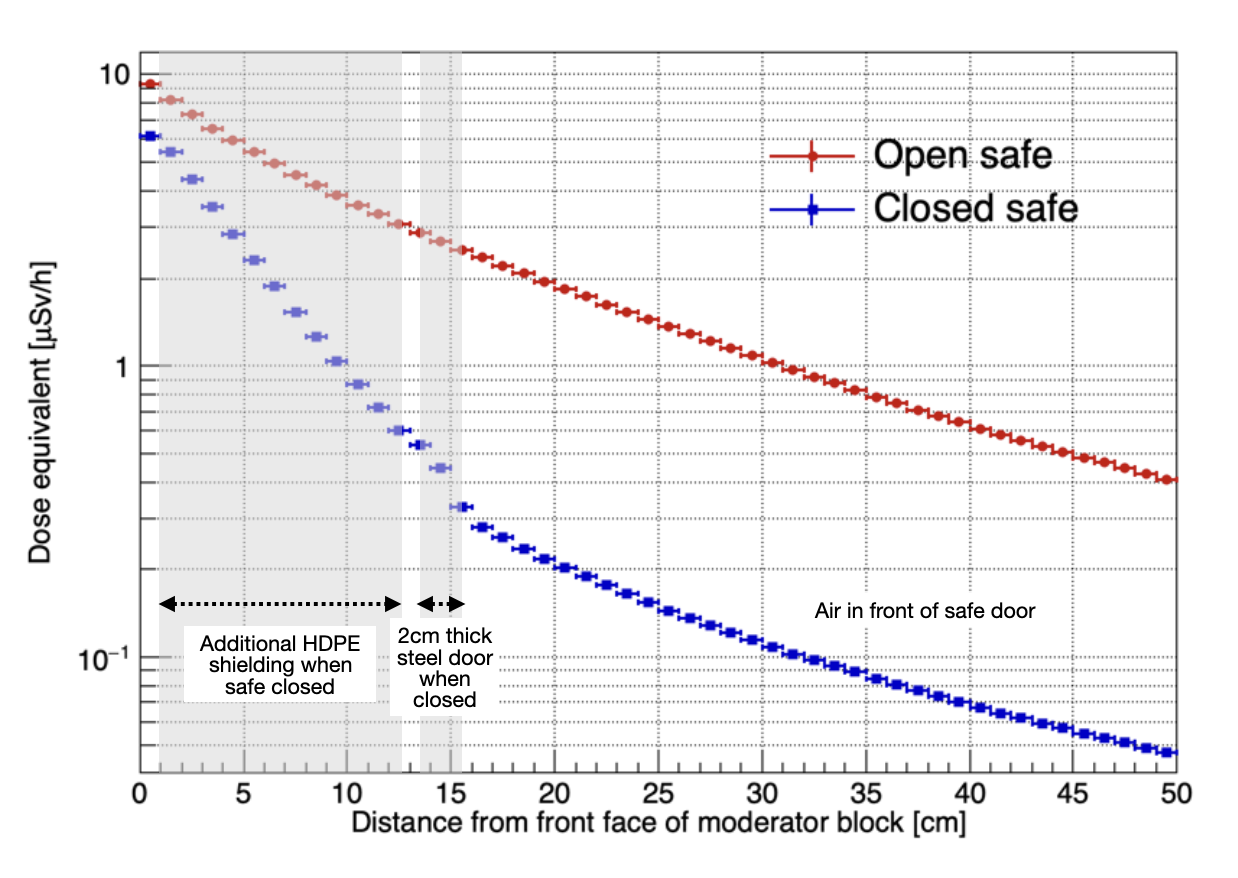}
    \caption{FLUKA simulation of equivalent dose rates close to the facility in the safe-opened and closed configurations.}
    \label{fig:eqdose2}
\end{figure}

The radiological impact of the facility was assessed using FLUKA simulations for
both open and closed safe configurations. Dose-equivalent rates were scored
using \texttt{USRBIN}. The scoring volume extended from the front face of the
moderator assembly, over an area of 8\,cm\,$\times$\,8\,cm, and was divided into
1\,cm bins along the forward direction extending approximately 35\,cm in front
of the steel safe door. The resulting dose-equivalent profiles are shown in
Figure~\ref{fig:eqdose2}.

For the closed-safe configuration, the combined effect of the HDPE shielding and
steel enclosure is clearly observed. The predicted dose-equivalent rate at the
safe door is approximately 0.3\,$\mu$Sv/h, falling to approximately
0.08\,$\mu$Sv/h at 20\,cm in front of the door. These values are comparable to,
or below, typical UK natural background levels, approximately
0.3\,$\mu$Sv/h. Measurements performed close to the facility using a Bonner
sphere dosimeter and a standard dose-rate monitor did not show readings above
background, consistent with the simulated prediction of a low external
radiological impact in the closed configuration.

The open-safe configuration represents the short period during which a detector
is inserted or removed from the moderator assembly. In the simulation, detector
position~4 was left open without the corresponding shielding HDPE filler, representing a
conservative configuration from a radiological perspective. Higher
dose-equivalent rates are predicted close to the source access region, with
values of order several \,$\mu$Sv/h in the region where hands may be present
during detector positioning. However, such operations are brief, typically
lasting only a few seconds, and therefore correspond to a very small integrated
dose. This operational mode is compatible with the intended use of the facility,
provided normal local radiation protection procedures are followed.

Overall, the simulations indicate that the facility has a low radiological
impact during normal operation. In the closed configuration, external
dose-equivalent rates remain at or below typical natural background levels,
while the open configuration is limited to short-duration detector handling
operations. Together with the permanent housing of the AmBe source within the
moderator assembly, this supports safe operation of the facility within a
standard university laboratory environment.

\section{Conclusions}

A compact AmBe-based neutron irradiation facility has been designed, commissioned and characterised for neutron detector research and development in a typical university laboratory environment. 
A modular HDPE moderator assembly provides defined detector positions with neutron fields ranging from fast-neutron dominated to strongly moderated conditions.

Detailed FLUKA simulations were used to model neutron and gamma transport within the facility and to predict the response of a commercial diamond neutron detector. 
Comparison between simulated and measured deposited-energy spectra at representative fast and thermal-neutron dominated positions shows good agreement above approximately 300--400\,keV in both spectral shape and absolute normalisation, demonstrating that the radiation fields within the facility are well understood and accurately described by the Monte Carlo model.

The simulations also provide insight into the detector response, including recoil processes, neutron capture in the $^6$LiF conversion layer, and secondary charged-particle production in surrounding materials. 
The detector assembly is shown to modify the local neutron field, with a reduction in thermal neutron flux at the sensor location of up to a factor of three. 
This highlights the importance of including realistic detector geometry, support structures and packaging materials when modelling and benchmarking detector response.

Unlike conventional benchmarking approaches based on activation foils, this methodology validates the facility directly in terms of detector response across the neutron and gamma radiation field, 
providing a continuous, measurement based characterisation that is particularly well suited to neutron detector research and development. 
The ability to interpret the response in terms of specific nuclear interaction channels, including $^6$Li(n,$\alpha$)$^3$H capture and recoil processes in carbon, 
further demonstrates that the radiation environment is both well characterised.

The use of a mixed neutron--gamma field further enhances the relevance of the facility, 
as it reflects realistic radiation environments in which gamma backgrounds accompany neutron fluxes. 
This enables simultaneous evaluation of neutron sensitivity and gamma response, 
which is essential for practical detector development.

The radiological assessment shows that the facility can be operated with low external dose rates and without routine handling of the radioactive source, 
which remains permanently housed within the moderator assembly. 
This supports safe operation consistent with the ALARP principle and reinforces its suitability for routine laboratory use.

Overall, the combined simulation and experimental results demonstrate that the facility provides a reliable and accessible platform for neutron detector development and benchmarking, 
and establishes a well-characterised reference environment for future detector studies.

\section*{Acknowledgements}
The authors thank Peter\,R.\,Hobson and Paul\,S.\,Miyagawa for their critical reviews of the manuscript and for constructive feedback that improved the clarity and robustness of this work. 
The authors also thank Dominic Howgill for technical support and for the design and fabrication of the detector holders, which enabled reproducible detector positioning within the facility. 
The authors further acknowledge the university technical workshop support of Jack Parr and Douglas Thomson for machining the HDPE components.
The authors also thank the university radiation protection team, Mark Ariyanayagam, Alan\,J.\,Drew and Saqib Qureshi, for support and guidance and ensuring regulatory compliance.
This work was supported by grants from the UK Science and Technology Facilities Council (STFC). 

\bibliographystyle{elsarticle-num}
\bibliography{references_v2}

\end{document}